\documentclass[preprintnumbers,eqsecnum,aps,prd,epsf,showpacs,nofootinbib,
twocolumn]{revtex4}
\usepackage{latexsym,amsmath,amssymb}
\usepackage[dvips]{graphicx,color}
\usepackage{dcolumn,bm}    
\usepackage{subfigure}  
\usepackage{color}
\input{colordvi.tex}

\newcommand{\ma}[1]{{\mathrm{#1}}}
\newcommand{\calN}{{\cal N}}
\newcommand{\calM}{{\cal M}}
\newcommand{\calO}{{\cal O}}

\newcommand{\cale}{{\cal E}}

\newcommand{\pa}{{\partial}}
\newcommand{\na}{{\nabla}}
\newcommand{\calg}{{\cal G}}
\newcommand{\frR}{{\frak R}}

\begin{document}
\thispagestyle{empty}
\title{Gradient expansion formalism for nonlinear 
superhorizon perturbations}

\preprint{}
\author{Yu-ichi Takamizu$^{1}$}
\email{takamizu_at_ccs.tsukuba.ac.jp}

\affiliation{
$^{1}$ Center for Computational Sciences, University of Tsukuba, 
1-1-1 Tennoudai, Ibaragi 305-8577 Japan 
}
\date{\today}

\begin{abstract}
We develop a theory of nonlinear cosmological perturbations on
 superhorizon scales where a characteristic length scale of perturbations is longer than the Hubble radius, 
in general theoretical frameworks. Our formalism is 
based on the spatial gradient expansion approach 
by adopting the ADM decomposition. Nonlinear superhorizon perturbation including both scalar (curvature perturbation) and tensor (gravitational waves) modes can be dealt with valid up to a second-order in the expansion. 
First we will review the formalism for a standard general relativity (GR) gravity plus a general kinetic single scalar (k-inflation) with a general form of the potential in the context of inflationary cosmology. That is the basic overview of our procedure.  Then it can be extended to more general framework, that is (1) 
beyond k-inflation (Galileon inflation), (2) a multi-component scalar field 
 with a general kinetic term and a general form of the potential and also 
(3) beyond Einstein gravity (general scalar-tensor theory), which can lead to 
several kinds of modified gravity. These theories are motivated not only 
inflation, but also the topic of dark energy. 
We provide a formalism to obtain the solution and construct nonlinear curvature perturbation in such general theoretical situation and it can be applied to the calculation of the superhorizon evolution 
 of a primordial non-Gaussianity beyond the so-called $\delta N$ formalism, showing fully nonlinear interaction of both scalar and tensor modes.    
\end{abstract}
\maketitle

\section{Introduction}
Cosmological nonlinear perturbation on superhorizon scale plays a key role to 
investigate evolution of primordial perturbation including scalar (curvature) perturbation and tensor (gravitational waves) perturbation. Especially, non-Gaussianity of curvature perturbations is one of the most powerful tool to distinguish different models of inflation (see Ref. \cite{CQG-focus-NG} and references therein). And also, in the superhorizon region where the characteristic length scale is longer than the Hubble radius, one may consider a classification of quantum fluctuations stretched over from subhorizon scale, however, this fundamental  process is unknown. In order to make clear this physics, 
it is important to investigate evolution of superhorizon perturbation, especially, the classical evolution equation followed by such superhorizon perturbation in the view point of link to a quantum equation. 
We develop nonlinear cosmological perturbation by adopting 
a gradient spatial expansion \cite{Salopek:1990jq}, 
different from the standard second-order 
perturbation theory \cite{Maldacena:2002vr}.  On superhorizon scale, such approach is a powerful tool to allow us to calculate {\it full} nonlinear effect 
in terms of standard perturbation language. The zeroth-order in gradient expansion is equivalent to the formalism called $\delta N$ formalism \cite{Starobinsky:1986fxa,Nambu:1994hu,Sasaki:1995aw,Sasaki:1998ug}, so our next-leading order in expansion can be called {\it beyond} $\delta N$ formalism \cite{Tanaka:2006zp,Takamizu:2008ra, Takamizu:2010xy, Takamizu:2010je, Naruko:2012fe, Takamizu:2013gy, Takamizu:2013wja, Takamizu:2018}. It can 
contain higher order contributions, related to a 
violation of slow-roll condition. 

The recent observational data of PLANCK satellite in 2015 
\cite{Ade:2015ava} 
gave us 
detailed observational data of Cosmic Microwave Background and the fact that 
non-Gaussianity of primordial curvature perturbation is very small at the local 
type, which is predicted by using $\delta N$ formalism. Indeed $\delta N$ formalism just leads to  {\it constant} contribution which only related to the 
local type of non-Gaussianity, but our beyond $\delta N$ obtain a {\it time evolution} and general kind of non-Gaussianity. We hope that the future precision detection of non-Gaussianity may actually be expected  a frequency dependence. 
Moreover, primordial gravitational waves can be expected to be detected in the near future and general prediction of tensor perturbation in nonlinear 
cosmological perturbation in a general theoretical setup, 
such as a modified gravity needs. Thus, to evaluate such superhorizon perturbations, it is necessary to develop a nonlinear theory of 
cosmological perturbations valid up through the next-leading order in the gradient expansion. 

In this paper, we will overview our basic procedure 
in Sec. \ref{single-k} as a proto type of formalism for a single general 
kinetic inflation (k-inflation) \cite{Takamizu:2010xy}. 
Then we discuss the extension of our formalism for beyond k-inflation, 
that is Generalized Galileon (G-inflation) \cite{Takamizu:2013gy} in Sec. III. 
In Sec. IV and Sec. V, we develop extensions of our formalism for multi-scalar 
case \cite{Naruko:2012fe} and beyond GR gravity plus a single scalar, 
namely the most general scalar-tensor theory (beyond Horndeski theory), respectively. 
 Section VI is devoted to the 
conclusion. 
\section{Single-field case}
\label{single-k}
In this section, the model of non-canonical single scalar field is a 
good example as a basic review of our formalism following 
\cite{Takamizu:2008ra,Takamizu:2010xy}. Then 
we consider GR gravity plus a 
general kinetic single scalar field
described by the Lagrangian density of the form 
\begin{align}
 {\cal L}=\sqrt{-g}\left[ {^{(4)}R\over 16\pi G_N}+P(X,\phi)\right]\,,
\end{align}
where $^{(4)} R$ is the four-dimensional Ricci scalar and 
$X:=-g^{\mu\nu}\partial_{\mu}\phi\partial_{\nu}\phi$/2. 
Note that we do not 
assume the explicit forms of both kinetic term and its 
potential, that can be given as arbitrary function of $P(X,\phi)$. 
Hereafter we will adopt units such that $8\pi G_N=1$. 

We introduce a small  expansion parameter: $\epsilon\equiv 1/(HL)$, which is the ratio of the Hubble length scale $1/H$ to the characteristic length scale of perturbations $L$ and the order in expansion can be expressed as 
$\calO(\epsilon^2)$. 

First of all, we show the main result in our formula for the 
nonlinear curvature perturbation: ${\cal R}_c^{\rm NL}$, 
\vspace{-0.3cm}
\begin{eqnarray}
\pa_\tau^2 {{\cal R}_c^{\rm NL}}+2 {\pa_\tau z\over z} 
\pa_\tau {{\cal R}_c^{\rm NL}}+{c_s^2\over 4} {\cal K}^{(2)}[\,
{\cal R}_c^{\rm NL}\,]=\calO(\epsilon^4)\,,
\label{O29-05_eq: basic eq for NL}
\end{eqnarray}
which shows two full-nonlinear effects; 
\vspace{-0.2cm}
\begin{enumerate}
\item Nonlinear variable: ${\cal R}_c^{\rm NL}$ including full-nonlinear curvature perturbation, $\delta N$
\vspace{-0.2cm}
\item Source term: ${\cal K}^{(2)}[{\cal R}_c^{\rm NL}]$ is a nonlinear function of curvature perturbations. 
\end{enumerate}
In (\ref{O29-05_eq: basic eq for NL}), $\tau$ denotes a 
conformal time and $z$ is a well-known Mukhanov-Sasaki variable: 
\begin{align}
z={a\over H}\sqrt{\rho+P \over c_s^2}\,,
\label{Mukha-Sasaki-v}
\end{align}
where $\rho$ and $P$ denote energy density and pressure of a scalar field, 
respectively with a speed of sound for perturbation: $c_s^2$ whose explicit 
definition will be shown later (\ref{speed-sound}). 
The definition of ${\cal R}_c^{\rm NL}$  
will be also seen later, in (\ref{O29-05_def0: nonlinear variable zeta}) 
and the source term 
${\cal K}^{(2)}[\gamma]$ 
is the Ricci scalar of the metric $\gamma$, respectively, whose 
explicit form will be shown in (\ref{def: K2}). 
Of course, 
in the linear limit, it can be reduced to the well-known equation
for the curvature perturbation on comoving hypersurfaces; $
\pa_\tau^2 {{\cal R}^{\rm Lin}_c}+2{\pa_\tau z\over z} \pa_\tau 
{{\cal R}^{\rm Lin}_c}
-c_s^2\,\Delta[{\cal R}^{\rm Lin}_c]=0$, with 
the Laplacian $\Delta\equiv\nabla^2$. 

We will briefly summarize our formula and 
show the above results in the following. 
We adopt the Arnowitt-Deser-Misner (ADM) 
decomposition and employ the gradient expansion. 
In the ADM decomposition, the metric is expressed as 
\begin{align}
ds^2 = - \alpha^2 dt^2 + g_{ij}(dx^i+\beta^idt)(dx^j+\beta^jdt)\,,
\end{align}
where $\alpha$ is the lapse function, $\beta^i$ is the shift vector and
Latin indices run over $1, 2, 3$. 
We introduce the extrinsic 
curvature $K_{ij}$ defined by  
\begin{align}
K_{ij} =
  \frac{1}{2\alpha}\left(\partial_t g_{ij}-\na_i\beta_j-\na_j\beta_i\right)\,,
\end{align}
where $\na$ is the covariant derivative compatible with the spatial metric
$g_{ij}$. As a result, 
the basic equations are reduced to 
the first-order equations for the dynamical variables $(g_{ij}$,$K_{ij})$,  with the two constraint equations. 
We further 
decompose them as 
\begin{align}
 &g_{ij} =  a^2(t) e^{2 \zeta}\gamma_{ij}\,,\notag\\
 &K_{ij}  =  
  a^2(t) e^{2 \zeta}\left(\frac{1}{3}K{\gamma}_{ij}
	    +{A}_{ij}\right)\,,
\end{align}
where $a(t)$ is the scale factor of the Universe for the background spacetime. 
$\gamma_{ij}$ is an unit-determinant metric ${\rm det} [{\gamma}_{ij}]=1$ and 
$A_{ij}$ is the traceless part of the extrinsic curvature. 
And also, $K$ is defined by $K\equiv \gamma^{ij} K_{ij}$. 
We choose a spatial gauge choice as 
\begin{align}
\beta^i=0\,.
\label{assum:beta-zero}
\end{align}
That simplifies the basic equations because it means naively ignoring any 
vector modes. Of course, one can take into account the condition 
$\beta^i\neq 0$. In that case, one can obtain vector modes as referring 
\cite{Takamizu:2008ra}. Hereafter we will take this simple spatial gauge choice (\ref{assum:beta-zero}) 
throughout this paper. 
In this gauge choice, we obtain evolution equations for curvature 
perturbation $\zeta$ and tensor perturbation $\gamma_{ij}$ as 
\begin{align}
 \pa_\perp \zeta
 &= -\frac{H}{\alpha} 
 + \frac{K}{3}\,, 
\label{basic-zeta} \\
 \pa_\perp \gamma_{i j}
 &=  2 A_{i j}\,,
\label{basic-gamma}
\end{align}
where $\pa_\perp\equiv \pa_t/\alpha$ and $H$ is the Hubble parameter defined 
by $H(t)\equiv \dot{a}(t)/a(t)$ for the background Friedmann-Lemaitre-Robertson-Walker (FLRW) spacetime. Hereafter a dot denotes represents differentiation with respect to $t$. 
They were derived from the definitions of $K$ and $K_{ij}$ given above. 
And also, two dynamical equations for $(K, A_{ij})$ can be obtained by varying the Lagrangian with respect to $\gamma_{ij}$, that corresponds to trace part and traceless part as 
\begin{align}
 \pa_\perp K &= -\frac{1}{3} K^2 - A_{i j} A^{i j}
 + \frac{1}{a^2 e^{2\zeta} \alpha} \Bigl(  D^2 \alpha + D_i \alpha D^i \zeta 
\Bigr)\notag\\
 & - \frac{1}{2} \left( S + E \right), \\
 \pa_\perp A_{i j} &=- K A_{i j} + 2 A_i{}^k A_{k j}\notag\\
 & - \frac{1}{a^2 e^{2\zeta}} \Bigl[ R_{i j}
 + D_i \psi D_j \zeta- D_i D_j \zeta \notag\\
 & - \frac{1}{\alpha} \Bigl( D_i D_j \alpha - D_i \alpha D_j \zeta 
 - D_j \zeta D_i \alpha \Bigr) \Bigr]^{TF} + S_{i j} ,
\end{align}
where $D$ is the covariant derivative compatible with $\gamma_{ij}$, 
$D^2 \equiv \gamma^{ij} D_i D_j$, $R_{ij} \equiv R_{ij}[\gamma]$, that is 
the Ricci tensor of the spatial metric $\gamma_{ij}$, 
and $[\cdot]^{\rm TF}$ means the 
trace-free operator, which is defined by $Q^{\rm TF}_{ij}\equiv Q_{ij}-
\gamma_{ij} \gamma^{kl}Q_{kl}/3$.  
$\gamma^{ij}$ is the inverse matrix of $\gamma_{ij}$ 
 and the index of $A_{i j}$ is raised by $\gamma^{i j}$. 
And also, the matter field part can be 
given by the energy-momentum tensor $T_{\mu\nu}$ as 
$E\equiv T_{00}/\alpha^2$ and $T_{ij}=a^2(t) e^{2\zeta} (S \gamma_{ij}/3+
S_{ij})$ with $S\equiv \gamma^{ij}T_{ij}$. 

Varying $\alpha$ and $\beta^i$ 
gives two constraints called {\it Hamiltonian} and {\it Momentum} constraint equations, respectively, which are 
\begin{align}
 &\frac{1}{a^2 e^{2\zeta}} \Bigl[ R - (4 D^2 \zeta+2D^i \zeta  D_i \zeta ) 
\Bigr]
 + \frac{2}{3} K^2 - A_{i j} A^{i j} = 2E, \\ 
 &\frac{2}{3} \pa_i K - e^{-3\zeta}D_j \Bigl( e^{3\zeta} A^j{}_i \Bigr) =J_i, 
\end{align}
where $R \equiv R[\gamma]$ is the Ricci scalar of the 
 normalized spatial metric $\gamma_{ij}$ and $J_i=-P_X\pa_i \phi$. 

The equation of motion for $\phi$ is given by 
\begin{equation}
 \frac{2}{\sqrt{-g}}\partial_{\mu}\left(\sqrt{-g}P_X\partial^{\mu}\phi\right)
  + P_{\phi} = 0,
  \label{eqn:EOM-phi}
\end{equation}
where the subscripts $X$ and $\phi$ represent derivative with respect to
$X$ and $\phi$, respectively. 
\subsection{Gradient expansion approach}
Next, we will employ the gradient expansion. In this approach 
we introduce a flat FLRW universe 
($a(t)$, $\phi_0(t)$) as a background. As discussed, 
we consider the perturbations on superhorizon scales, therefore 
we consider $\epsilon= 1/(HL)=k/(aH)$ as a small expansion 
parameter and systematically expand equations by $\epsilon$. Spatial derivative acting on a perturbation raises the order by $\epsilon$ as $\partial_i Q=\calO(\epsilon) Q$. We attach the superscript $(m)$ to a quantity of 
$\calO(\epsilon^m)$ throughout this paper. 

It is natural to assume the condition for the gradient expansion; $\partial_t{\gamma}_{ij} = \calO(\epsilon)$ since the FLRW universe is recovered as 
background for $\epsilon\to 0$. Moreover, we assume the stronger 
condition: 
\begin{align}
\partial_t{\gamma}_{ij} = \calO(\epsilon^2)\,.
\label{stronger condition for gamma}
\end{align}
This corresponds to assuming the absence of any decaying modes
at the leading-order in the expansion. This is 
justified in taking inflationary acceleration of the universe. In fact, under the condition: $\partial_t{\gamma}_{ij} = \calO(\epsilon)$, we obtain 
the general solution $\propto a^{-3}$ in \cite{Takamizu:2018} and 
also in \cite{Frusciante:2013haa}, 
so inflation can wash away such decaying mode soon. 
In order to solve the above basic equations, one has to fix the gauge 
condition. For the case of single scalar, the most convenient choice of the temporal coordinate is such that the expansion $K$ is uniform and takes the form: 
\begin{align}
K(t,x^i)=3 H(t)\,,
\end{align}
called {\it uniform expansion gauge}. Adopting this gauge choice, 
the basic equation reduces simply to 
\begin{align}
\pa_t \zeta= H(\alpha -1)=: H \delta \alpha(t,x^i)\,.
\label{pa_zeta-alpha}
\end{align}
It means that 
the time evolution of the curvature perturbation caused by the inhomogeneous 
part of the lapse function $\delta \alpha$ only, relating to the non-adiabatic 
perturbation on superhorizon scale. Hereafter $\delta$ denotes a fluctuation for any quantity as $\delta Q \equiv Q-Q_0$, where the subscript $0$ denotes the background. 
With the above choice of gauge, the general solution valid up to 
$\calO(\epsilon^2)$ can be obtained in \cite{Takamizu:2008ra}. 

When we focus on a contribution arising from the scalar-type 
perturbations, we may choose the gauge in which ${\gamma}_{ij}$ approaches the flat metric as $\gamma_{ij}\to \delta_{ij}$ for $t\to \infty$, when the epoch 
close to the end of inflation. 
For the purpose of construct nonlinear curvature perturbation,  we take the {\it comoving slicing}: $
\delta \phi_c(t,x^i)=0$. 
Let us use the subscript $K$ and $c$ to indicate the quantity in the 
uniform expansion$(K)$ and comoving gauge. So we had obtained the general solution, that is attached to the subscript $K$. 
By using a transformation from uniform $K$ gauge to the comoving gauge, we obtain the curvature perturbation in the comoving gauge: $\zeta_c=\zeta_K-H
\delta \phi_K/\dot{\phi} +\calO(\epsilon^3)$. 

Now we turn to the problem of properly defining a nonlinear
curvature perturbation to $\calO(\epsilon^2)$ accuracy by using the general 
solution above. 
Hereafter we will use the expression ${\cal R}_c$ on 
comoving slices to denote it. 
Let us consider 
the linear curvature perturbation which is  
given as $
{\cal R}^{\rm Lin}=\left(H^{\rm Lin}_L+{H^{\rm Lin}_T\over 3}\right)Y$, 
where, following the notation in \cite{Kodama:1985bj},
the spatial metric in the linear limit is expressed as $
\gamma_{ij}=a^2(\delta_{ij}
+2 H^{\rm Lin}_LY \delta_{ij}+2 H^{\rm Lin}_T Y_{ij})$.  
These expressions in the linear theory 
correspond to the metric components in our
notation as ${\zeta}=
 H^{\rm Lin}_L Y$ and 
${\gamma}_{ij} = \delta_{ij} + 2H^{\rm Lin}_T Y_{ij}$. 
Notice that the variable ${\zeta}_c$ reduces to 
${\cal R}^{\rm Lin}_c$ at leading-order in the gradient expansion, 
but not at second-order and it will be also similar to the nonlinear theory. 
Thus to define a nonlinear generalization of the linear curvature
perturbation, we need nonlinear generalizations
of $H_LY$ and $H_TY$. Our nonlinear ${\zeta}$ is an
apparent natural generalization of $H^{\rm Lin}_LY$ as $H_LY=\zeta$. 
As for $H_TY$, however, the generalization is non-trivial.
It corresponds to the $\calO(\epsilon^2)$ part of $\gamma_{ij}$. 
As shown in \cite{Takamizu:2010xy}, it can be 
done by introducing
the inverse Laplacian operator $\Delta^{-1}$ on the flat background and 
we defined the nonlinear generalization of $H_TY$ as
\begin{eqnarray}
H_TY=\chi \equiv
-\frac{3}{4}\Delta^{-1}\left[\partial^i e^{-3 \zeta^{(0)}}
\partial^j e^{3 \zeta^{(0)}}(\ln {\gamma})_{ij} \right].
\label{O29-05_def: nonlinear HT0}
\end{eqnarray}
With these definitions of $H_LY$ and $H_TY$, we can define the nonlinear 
curvature perturbation valid up through $\calO(\epsilon^2)$ as
\vspace{-0.5cm}
\begin{eqnarray}
{\cal R}^{\rm NL}_c\,\equiv\, {\zeta}_c\,+\,{\chi_c\over 3}\,.
\label{O29-05_def0: nonlinear variable zeta}
\end{eqnarray}
It is easy to show that this nonlinear quantity can be reduced to 
${\cal R}^{\rm Lin}_c$ in the linear limit. 
As clear from (\ref{O29-05_def: nonlinear HT0}), finding $H_TY$ generally
requires a spatially non-local operation, however, 
in the comoving slicing with the asymptotic condition on the spatial 
coordinates, we find it
is possible to obtain the explicit form of $H_TY$ 
without any non-local operation as seen in \cite{Takamizu:2010xy}. 
Finally, we can derive 
a nonlinear second-order differential equation that 
${\cal R}_c^{\rm NL}$ (\ref{O29-05_def0: nonlinear variable zeta}) satisfies at 
$\calO(\epsilon^2)$ accuracy by introducing the conformal time $\tau$, defined by 
$d\tau={dt/a(t)}$ and 
the Mukhanov-Sasaki variable (\ref{Mukha-Sasaki-v}) 
with a speed of sound $c_s^2$: 
\begin{align}
c_s^2={P_X\over P_X+2 P_{XX} X}\,.
\label{speed-sound}
\end{align} 
The result can be reduced to a simple equation of the form 
(\ref{O29-05_eq: basic eq for NL}) 
as a natural extension of the linear version: 
\begin{eqnarray}
\pa_\tau^2{{\cal R}_c^{\rm NL}}+2 {\pa_\tau z\over z} 
\pa_\tau{{\cal R}_c^{\rm NL}} +{c_s^2\over 4} {\cal K}^{(2)}[\,
{\cal R}_c^{\rm NL}\,]=\calO(\epsilon^4)\,,
\label{eq: basic eq for NL}
\end{eqnarray}
with 
\begin{align}
{\cal K}^{(2)}[\zeta^{(0)}]=\, 
{-2 (2 \Delta \zeta^{(0)}+\delta^{ij}\partial_i \zeta^{(0)}
\partial_j \zeta^{(0)})e^{-2 \zeta^{(0)}}}\,.
\label{def: K2}
\end{align}
Through up to the second-order in expansion, it can be evaluated 
as ${\cal K}^{(2)}[{\cal R}_c^{\rm NL}]={\cal K}^{(2)}[\zeta^{(0)}_c]+\calO (\epsilon^4)$ since ${\cal R}_c^{\rm NL}=\zeta_c+ \calO (\epsilon^2)$ with $\chi_c=\calO (\epsilon^2)$.

\subsection{Linear Theory valid up to $\calO(\epsilon^2)$}
To obtain the power spectrum, we will use the linear theory of the curvature perturbation in this subsection. 
The master equation: 
\begin{align}
\pa^2_{\tau}{{\cal R}^{\rm Lin}_c}+2{\pa_\tau z\over z} \pa_\tau {{\cal R}^{\rm Lin}_c}
-c_s^2\,\Delta[{\cal R}^{\rm Lin}_c]=0\,, 
\label{eq: basic eq for Lin}
\end{align}
has two independent solutions; conventionally 
called a growing mode and a decaying mode. We assume that the growing mode
is constant in time at leading order in the spatial gradient expansion as the assumption: (\ref{stronger condition for gamma}),  justified in an inflationary universe. 

As shown in \cite{Takamizu:2010xy}, 
the linear solution valid up to $\calO(\epsilon^2)$ can be 
obtained as
\begin{eqnarray}
{\cal R}_{c,\bm{k}}^{\rm Lin}(\tau)
&=&\Bigl[\tilde{\alpha}^{\rm Lin}_{\bm{k}}+
(1-\tilde{\alpha}^{\rm Lin}_{\bm{k}}
){\tilde{D}(\tau)\over \tilde{D}_*}\nonumber\\
&&-\left({\tilde{F}_* \over \tilde{D}_*}\tilde{D}(\tau)+
\tilde{F}(\tau)\right)k^2\Bigr]U^{(0)}_{\bm{k}}\,,
\label{sol: linear Rc}
\end{eqnarray}
where $U_{\bm k}^{(0)}$ denotes an integration constant 
and the integrals $\tilde{D}(\tau)$ and 
$\tilde{F}(\tau)$ have been given as 
\begin{eqnarray}
\tilde{D}(\tau) & = & 3{\cal H}(\tau_*) 
  \int_{\tau_*}^\tau d\tau' {z^2(\tau_*) \over  z^2 (\tau')},
  \nonumber\\
 \tilde{F}(\tau) & = &
   \int_{\tau_*}^\tau \frac{d\tau'}{z^2(\tau')}
  \int_{\tau_*}^{\tau'}z^2(\tau'')c_s^2(\tau'')d\tau''.
\label{int-tilDF}
\end{eqnarray}
Here $\tilde{D}_*=\tilde{D}(\tau_*), \tilde{F}_*=\tilde{F}
(\tau_*)$, $\tau_*$ and $\cal H$ denote an initial time of 
gradient expansion and the conformal Hubble parameter 
${\cal H}=d \ln a/d\tau$, respectively.
The integrals in (\ref{int-tilDF}) represent a decaying and growing mode solution, respectively.

Note that ${\cal R}^{\rm Lin}_{c,_{\bm{k}}}(\tau_*)=U^{(0)}_{\bm{k}}$ that is 
just a constant solution, 
while ${\cal R}^{\rm Lin}_{c,_{\bm{k}}}(0)=\tilde{\alpha}^{\rm Lin}_{\bm{k}}U^{(0)}_{\bm{k}}$.
Thus if the factor $|\tilde{\alpha}^{\rm Lin}_{\bm{k}}|$ is large, it represents 
an enhancement of the curvature perturbation on superhorizon scales
due the $\calO(\epsilon^2)$ effect \cite{Leach:2001zf}.


Here it is useful to consider an explicit
expression for $\tilde{\alpha}^{\rm Lin}_{\bm{k}}$ in terms of 
${\cal R}^{\rm Lin}_{c,{\bm{k}}}$ and its derivative at $\tau=\tau_*$.
The result is 
\begin{eqnarray}
\tilde{\alpha}^{\rm Lin}_{\bm{k}}=1+ {\tilde{D}_*\over 3 {\cal H}_*}
\frac{{\pa_\tau{\cal R}^{\rm Lin}_{c,_{\bm{k}}}}(\tau_*)}{{\cal R}^{\rm Lin}_{c,_{\bm{k}}}(\tau_*)}
-k^2 \tilde{F}_*+\calO(k^4).
\end{eqnarray}

In order to relate our calculation with the standard formula for
the curvature perturbation in linear theory, we introduce 
$\tau_k$ (or $t_k$) which denotes the time at which the comoving wavenumber
has crossed the Hubble horizon,
\begin{equation}
\tau_k=-{r\over k}\,;\quad 0<r\ll 1 \,.
\end{equation}
The power spectrum at the horizon crossing time is given by 
%
\begin{align}
 &\langle {\cal R}^{\rm Lin}_{c,\bm{k}}(\tau_k)
  {\cal R}^{\rm Lin}_{c,\bm{k}'}(\tau_{k'})
  \rangle
  = (2\pi)^3P_{{\cal R}}(k)\delta^3(\bm{k}+\bm{k}'), \nonumber\\
  &P_{{\cal R}}(k) = 
  \left|{\cal R}^{\rm Lin}_{c,\bm{k}}(\tau_k)\right|^2\,.
  \label{eq: power spectrum for R}
\end{align}

By inverting ${\cal R}^{\rm Lin}_{c,\bm{k}}$ in terms of $U^{(0)}_{\bm{k}}$ as 
shown in \cite{Takamizu:2010xy}, we can show the final value
of the linear curvature perturbation as 
\begin{equation}
{\cal R}^{\rm Lin}_{c,\bm{k}}(0)=
 \tilde{\alpha}^{\rm Lin}_{\bm{k}}U^{(0)}_{\bm{k}} = 
  {\alpha}^{\rm Lin}_{\bm{k}}
  {\cal R}^{\rm Lin}_{c,\bm{k}}(\tau_k)
  + \calO(k^4)\,,
  \label{eqn:alphau}
\end{equation}
where
%
\begin{equation}
 {\alpha}^{\rm Lin}_{\bm{k}}
  =
  1 + \alpha^{\cal R}{D}_k- k^2 {F}_k \,,
  \label{til-alpha-lin} 
\end{equation}
and 
\begin{eqnarray}
 \alpha^{\cal R} & = & 
  \frac{1}{3{\cal H}(\tau_k) }
   \,{\pa_\tau {\cal R^{\rm Lin}}_{c,\bm{k}}\over {\cal R}^{\rm Lin}_{c,\bm{k}}}\bigg|_{\tau=\tau_k},
  \nonumber\\
 {D}_k & = & 3{\cal H}(\tau_k) 
  \int_{\tau_k}^0d\tau' {z^2(\tau_k) \over  z^2 (\tau')},
  \nonumber\\
 {F}_k & = &
 \int_{\tau_k}^0\frac{d\tau'}{z^2(\tau')}
  \int_{\tau_k}^{\tau'}z^2(\tau'')c_s^2(\tau'')d\tau''.
\end{eqnarray}
The formula (\ref{eqn:alphau}) will 
be used in the next subsection.

The power spectrum at the final time is thus enhanced by the factor 
$|{\alpha}^{\rm Lin}_{\bm{k}}|^2$ as
%
\begin{equation}
 \langle {\cal R}^{\rm Lin}_{c,\bm{k}}(0)
  {\cal R}^{\rm Lin}_{c,\bm{k}''}(0)
  \rangle
  = (2\pi)^3  |{\alpha}^{\rm Lin}_{\bm{k}}|^2
  P_{{\cal R}}(k)\delta^3(\bm{k}+\bm{k}')\,. 
  \label{eqn:powerspectrum}
\end{equation}
\subsection{Nonlinear theory valid up to $\calO(\epsilon^2)$}
Using the linear solution of the curvature perturbation 
given by (\ref{sol: linear Rc}), here we can derive
the nonlinear solution by matching the two at $\tau=\tau_*$. 
The main purpose of the matching is to make it possible to analyze
superhorizon nonlinear evolution valid up to the second-order in
gradient expansion, starting from a solution in the linear theory. In
particular, we would like to evaluate the bispectrum induced by the
superhorizon nonlinear evolution. For this purpose, we need to
have full control over terms up not only to $\calO(\epsilon^2)$ but also
to $\calO(\delta^2)$, where we suppose that the linear solution is of order
$\calO(\delta)$. Therefore, the matching condition at $\tau=\tau_*$ should
be of the form 
%
\begin{eqnarray}
 {\cal R}^{\rm NL}_c(\tau_*)
  & = & {\cal R}^{\rm Lin}_c(\tau_*) + s_1(\tau_*)
  + \calO(\epsilon^4,\delta^3), \nonumber\\
 \pa_\tau \,{{\cal R}^{\rm NL}_c}(\tau_*)
  & = & \pa_\tau\,{{\cal R}^{\rm Lin}_c}(\tau_*)
  + s_2(\tau_*) + \calO(\epsilon^4,\delta^3)\,,
\end{eqnarray}
where $ s_{1}(\tau_*) = \calO(\delta^2)$ and $s_{2}(\tau_*) = \calO(\delta^2)$ 
are functions of $\tau_*$ and spatial coordinates. While the linear
solution ${\cal R}^{\rm Lin}_c(\tau)$ is considered as an input,
i.e., initial condition, the additional terms, $s_{1}(\tau_*)$ and
$s_2(\tau_*)$, are to be determined by the following condition. 
The terms of order $\calO(\delta^2)$ in 
     ${\cal R}^{\rm NL}_{c,\bm{k}}$ and 
     $\pa_\tau{{\cal R}^{\rm NL}_{c,\bm{k}}}$ should vanish at the horizon crossing when $\tau=\tau_k$. Note that $\tau_k<\tau_*$. 
In other words, $s_1(\tau_*)$ and $s_2(\tau_*)$ represent the
$\calO(\delta^2)$ part of ${\cal R}^{\rm NL}_c$ and 
$\pa_\tau {{\cal R}^{\rm NL}_c}$, respectively, generated during the period between
the horizon crossing time and the matching time. 

We have to omit the explicit way 
to determine the terms $s_1$ and $s_2$ for want of space, that was 
shown in \cite{Takamizu:2010xy}. 
As a result, using the linear solution of the curvature perturbation 
given by (\ref{sol: linear Rc}) we have the nonlinear
comoving curvature perturbation at the final time $\tau=0$ (or
$t=\infty$) given by 
\begin{eqnarray}
&&{\cal R}_{c,\bm{k}}^{\rm NL}(0)=
{\cal R}^{\rm Lin}_{c,\bm{k}} (\tau_k) 
-(1-{\alpha}^{\rm Lin}_{\bm{k}})
{\cal R}^{\rm Lin}_{c,\bm{k}} (\tau_k) \nonumber\\
&&
-\frac{1}{4}F_k
  {\tilde R}^{(2)}[{\cal R}^{\rm Lin}_c(\tau_k)]
 +\calO(\epsilon^4, \delta^3)\,,
\label{sol: infty Rreal}
\end{eqnarray}
where 
\begin{align}
{\tilde R}^{(2)}[\zeta^{(0)}]&\equiv 
4\Delta \zeta^{(0)}
+{\cal K}^{(2)}[\zeta^{(0)}]\notag\\
&=-2(\delta^{ij}\partial_i \zeta^{(0)}
\partial_j\zeta^0-4 \zeta^{(0)}\Delta \zeta^{(0)})+\calO((\zeta^{(0)})^3)\,,
\end{align}
which  denotes the nonlinear term derived from Ricci scalar ${\cal K}^{(2)}$ 
(\ref{def: K2}). 
The first term in (\ref{sol: infty Rreal}) 
corresponds to the result of the $\delta N$ formalism, that is 
a constant since we considered the system for a single scalar field, 
the second term is related to an enhancement on superhorizon scales
in linear theory, and the last term is the nonlinear effect which
may become important if $F_k$ is large. 

Here we can notice that 
in order to the final values of curvature 
perturbation both in linear (\ref{eqn:alphau}) and in nonlinear theory (\ref{sol: infty Rreal}), all one have to do is to estimate the same integrals shown in both theories as $D_K$ and $F_k$ in $\alpha^{\rm Lin}_{\bm{k}}$. 
The reason why is 
that the master equations (\ref{eq: basic eq for NL}) and (\ref{eq: basic eq for Lin})
for both theories have 
the same structures of evolution equation as described before. 

In 
\cite{Takamizu:2010xy}, 
we calculated non-Gaussianity of curvature perturbations for a model of 
sudden slope change of inflaton's potential. We also had 
studied several applications of our formalism in 
\cite{Takamizu:2010je,Takamizu:2013wja}, 
where we considered a model of temporal stopping inflaton and varying sound speed, respectively. Such effect is sensitive to the temporal violation of 
some kind of slow-roll conditions. So our formalism is good to evaluate 
time evolution of curvature perturbation in the physics for violation of slow-roll conditions. 
\section{Beyond kinetic-inflation}
In this section, we will review \cite{Takamizu:2013gy} and 
we go beyond kinetic inflation whose the Lagrangian density given as 
%
\begin{equation}
{\cal L}= \sqrt{-g}\Bigl[ { ^{(4)}R\over 2} + 
P(X,\phi)-G(X,\phi)\Box\phi\Bigr], 
\label{scalar-lag}
\end{equation}
where $G$ is also an arbitrary function of $\phi$ and $X$. 
Although the above action depends upon the second derivative of $\phi$ through
$\Box\phi =g^{\mu\nu}\na_\mu\na_\nu\phi$,
the resulting field equations for $\phi$ and $g_{\mu\nu}$ remain second order.
In this sense the above action gives rise to a more general single-field inflation model
than k-inflation, {\em i.e.}, Generalized Galileon inflation(G-inflation)~\cite{Kobayashi:2010cm}. The same scalar-field Lagrangian is used
in the context of dark energy and called kinetic gravity braiding~\cite{KGB}. 
In fact, the most general inflation model with second-order field equations
was proposed in~\cite{Kobayashi:2010cm} based on Horndeski's scalar-tensor
theory~\cite{Horndeski, GenG}.
However, here we focus on
the action~(\ref{scalar-lag}) which
belongs to a subclass of the most general single-field inflation model,
because it involves sufficiently new and interesting ingredients
while avoiding unwanted complexity. 

We can replace the following equation of 
motion with (\ref{eqn:EOM-phi}) for this generic single scalar field as 
\begin{equation}
\nabla_{\mu}\Bigl[(P_X-G_\phi-G_X \Box\phi)\na^\mu \phi-G_X 
\na^\mu X\Bigr]+P_\phi-G_\phi\Box\phi= 0.
\end{equation}

Taking same procedure in the previous section can allow us to 
obtain the nonlinear curvature perturbation in the 
comoving gauge satisfying the similar form of evolution equation as 
a second-order differential equation: 
\begin{eqnarray}
\pa_\tau^2 {{\frR}_c^{\rm NL}}+2 {\pa_\tau z\over z} 
\pa_\tau{{\frR}_c^{\rm NL}} +{c_s^2\over 4} \,{\cal K}^{(2)}[\,
{\frR}_c^{\rm NL}\,]=\calO(\epsilon^4)\,,
\label{basic eq for NL}
\end{eqnarray}
where ${\cal K}^{(2)}[\,{\frR}_c^{\rm NL}\,]$ is the Ricci scalar
of the metric $\delta_{ij}\exp\left(2\frR_c^{\rm NL}\right)$ 
in (\ref{def: K2}) and 
\begin{eqnarray}
z:={a\dot{\phi}\sqrt{\calg_G} \over \Theta_G}\,,
\label{def: variable-z}
\end{eqnarray}
with 
\begin{align}
&{\cal{G}}_G(t) := \cale_X-3\Theta_G (G_{X}\dot{\phi})_0,\\
&\Theta_G(t)  := H -(G_X X \dot{\phi})_0,\label{def:cale-X}\\
&\cale_X (t) := \Bigl[P_X+2 X P_{XX}+9H G_X \dot{\phi}\notag\\
&+6 H G_{XX} 
\dot{\phi}_0 -2 G_\phi-2 X G_{\phi X}\Bigr]_0.\label{def:cale-X}
\end{align}
Here $c_s^2$ is the sound speed squared
of the scalar fluctuations defined as
\begin{align}
c^2_s:= {{\cal F}_G (t)\over \calg_G(t)},\quad
{\cal F}_G (t):={1\over X_0} (-\pa_t \Theta_G +\Theta_G G_X X
\dot{\phi})_0. 
\end{align}
This is a generalization of familiar ``$z$''
in the Mukhanov-Sasaki equation~\cite{Mukhanov:1990me},
and reduces indeed to $a\sqrt{(\rho+P)}/H c_s$
in the case of k-inflation as shown in (\ref{Mukha-Sasaki-v}). 

We have introduced an appropriately defined variable for 
the nonlinear curvature perturbation in the comoving gauge ($\delta\phi=0$), 
${\frR}_c^{\rm NL}$. It is a combination of $\zeta$ and scalar mode $\chi$ 
from an unit spatial metric $\gamma_{ij}$, which is given by 
\begin{align}
{\frR}^{\rm NL}\equiv \  {\psi}\,+\,{\chi\over 3}\,.
\label{def0: nonlinear variable zeta1}
\end{align}
with 
\begin{eqnarray}
\chi \equiv\ 
-\frac{3}{4}\Delta^{-1}\left[\partial^i e^{-3\psi}
\partial^j e^{3\psi}({\gamma}_{ij}-\delta_{ij}) \right]\,.
\label{def: nonlinear HT1}
\end{eqnarray}
The definition of (\ref{def: nonlinear HT1}) is same as 
(\ref{O29-05_def: nonlinear HT0}). The evolution equation for 
$\gamma_{ij}$ is obtained as 
\begin{align}
\pa_\tau^2 \gamma_{ij}+2{\cal H}\pa_\tau \gamma_{ij}+ 2 {\cal 
F}_{ij}^{(2)}[\gamma]=\calO(\epsilon^4)\,,
\label{evolution gamma GR}
\end{align}
where the explicit form of ${\cal F}_{ij^{(2)}}$ will be shown in (\ref{source-interaction}). 
That is worth comparing with the later result of 
(\ref{evolution eq for gamma GW}) for the beyond GR theory shown in 
Sec. V, where it is noticed 
that this equation is a key basic equation for gravitational waves. 

Upon linearization, the variable: ${\frR}^{\rm NL}$
reduces to the previously defined linear curvature perturbation
${\cal R}_c^{\rm Lin}$ on uniform $\phi$ hypersurfaces.
Then, it has been shown that ${\frR}_c^{\rm NL}$
satisfies a nonlinear second-order differential equation~(\ref{basic eq for NL}),
which is a natural extension of the linear perturbation equation for ${\cal R}_c^{\rm Lin}$ in (\ref{eq: basic eq for Lin}) and  ${\cal R}_c^{\rm NL}$ in (\ref{eq: basic eq for NL}). 
\section{Multi-field case}
In this section, we will review multi-field case following 
\cite{Naruko:2012fe}. 
We consider Einstein gravity plus a 
multi-component scalar field 
 described by Lagrangian density of the form 
\begin{align}
{\cal L}=& \sqrt{-g} \left[ { ^{(4)}R\over 2}+
P(X^{IJ},\phi^K)\right]\,,\notag\\
&X^{IJ} \equiv - g^{\mu \nu} \pa_\mu \phi^I \pa_\nu \phi^J/2\,,
\end{align}
where $I,J$ and $K$ run over 
$1,2,\ldots, \calM$ denoting $\calM$-components scalar field. 
Note that we do not assume the explicit form of 
both kinetic terms 
and their potentials, that can be given as arbitrary function of 
$P(X^{IJ},\phi^K)$. 
Notice that it can not be a perfect fluid form, that is different 
from a single scalar system and we can replace the following equation of 
motion with (\ref{eqn:EOM-phi}) for multi-scalar field as 
\begin{align}
 &\pa_\perp \Bigl( P_{( I J )} \pa_\perp \phi^J \Bigr)
 + K P_{( I J )} \pa_\perp \phi^J\notag\\
 &- \frac{1}{\alpha a^3 e^{3 \zeta}} \pa_i \Bigl( \alpha a e^\zeta P_{( I J )}
 \gamma^{i j} \pa_j \phi^J \Bigr) - \frac{1}{2} P_I=0.
\label{EOM:scalar}
\end{align} 
This is $\calM$-coupled second order differential equations. 

We introduce the proper time $\eta$ and its definition is 
\begin{gather}
 \eta (t, x^i) \equiv \int_{x^i = const.} dt ~ \alpha (t, x^i)\,.
\label{conformaltime}
\end{gather}
In terms of $\eta$, the expression of $K$ at leading order 
 in gradient expansion with a spatial gauge choice: $\beta^i=0$, 
\begin{align}
K = \frac{1}{\alpha}
 \frac{\pa_t ( a^3 \zeta^6 )}{a^3 \zeta^6} 
 =  3 \frac{\pa_\eta ( a \zeta^2 )}{a \zeta^2}\,.
\end{align}
If we associate $a \zeta^2$ and $\eta$ with $a$ and $t$ respectively,
 we can check the correspondence between $K$ and $3H$. 
Based on these facts, the structure of the above equations is same as
 that of background equations. Namely, given a background solution $
 \phi^I (t) |_\ma{background} = \phi^I_0 (t)$, one can 
construct the solution at leading order  in gradient expansion as 
$\phi^I (t, x^i) |_\ma{gradient} = \phi^I_0 (\eta)$. 

In the standard cosmological perturbation,  
 the $e$-folding number $N$ is often used as a time coordinate,
 which is defined by 
\begin{align}
N =  \int_t^{\infty} dt' H (t')\,.
\end{align} 
Note that it is the number of $e$-folding counted backward in time from a 
fixed final time $t=\infty$. 
 By replacing with $t$ with $\eta$ nd $H$ with $K/3$, we 
can generalize the e-folding number $\calN$ as 
 \begin{gather}
 \calN \equiv \frac{1}{3} \int_{t}^{t_0}\left. \alpha (t', x^i) \,
  K (t', x^i)\right|_{x^i=const.} d t'\,.
\label{O29-05_eq: def-calN}
 \end{gather}
If we also rewrite basic equation using $\calN$ as a time coordinate,
 we can again easily check that the structure of equations exactly
 coincide with that of background equations using $N$ as a time coordinate. 

As for a gauge choice, 
we consider the uniform $K$ slicing which is taken in multi field case. 
At the leading order in the gradient expansion,
 the evolution equation for the extrinsic curvature can be used in order to 
obtain the solution of lapse as 
\begin{gather}
 \alpha^{(0)} = - \frac{2 \dot{H} (t)}{E^{(0)} (t) + P^{(0)} (t, x^i)}
 + \calO (\epsilon^3),
\label{O29-05_eq: K,sol,alpha}
\end{gather}
We can express $\alpha^{(0)}$ as a function of $\eta$
 using $E^{(0)}$ and $P^{(0)}$. On the other hands, in a single-scalar system, 
it reads $\alpha^{(0)}=1$ and it is the main difference between 
single and multi case.  
Since we have assumed that we can solve the scalar field equation and
 express the scalar field as a function of $\eta$,
 we know the expressions of $E^{(0)}$ and $P^{(0)}$ as
 a function of $\eta$. 
This equation means the inhomogeneity of $\alpha$ is related with
 that of $P^{(0)}$. Therefore (\ref{O29-05_eq: K,sol,alpha}) can be schematically expressed as 
\begin{gather}
 \alpha = f \Bigl[ t, \, P (\eta) \Bigr]
 = f \Bigl[ t, \, P \Bigl( \int \alpha dt \Bigr) \Bigr] \,.
\end{gather}
It is clearly shown that it is almost impossible to solve
 this equation, at least in an analytical way. 

Then we have to solve the Einstein equations on another gauge, that is 
a uniform $\calN$ slice. 
From Eq (\ref{O29-05_eq: def-calN}),
 we can reread the condition of this slice as
\begin{gather}
 \alpha (t, x^i) K (t, x^i) =  3 H (t) ~~~ \Leftrightarrow ~~~ 
 \pa_t \zeta (t, x^i) = 0.
\end{gather}
It shows on this slice, curvature perturbation becomes constant which is expressed by a function of $x^i$ only. It is easy to obtain 
the solution in this gauge. First, we solve the Einstein equations on uniform
 $e-$folding number slice, the solution of scalar fields
 are given by the function of cosmic time or background $e-$folding number.
Next, we consider the gauge transformation from above slice
 to uniform expansion slice. Applying the derived gauge transformation rules to the solution on above slice,
 we can find out the solution on uniform expansion slice. 
\subsection{Nonlinear gauge transformation}
\label{app:gauge}
We derive the gauge transformation rules for the metric, 
 its derivative ($K$, $A_{i j}$) and the scalar field.
We consider a nonlinear gauge transformation from a coordinate system 
 with vanishing shift vector $\beta^i = 0$,
 to another coordinate system in which the new shift vector also vanishes,
 $\tilde{\beta}^i = 0$. We note that once the time slicing is changed,
 the shift vector appears in the new slicing in general.
So the spatial coordinates also need to be changed to 
 eliminate thus appeared shift vector. 
We use the background $e$-folding number $N$ as the time coordinate and
 define the temporal and spatial shift, $n$ and $L^i$, respectively, $
 \tilde{N} + \tilde{n} (\tilde{N}, \tilde{x}^i) = N,\ 
 \tilde{x}^i + \tilde{L}^i (\tilde{N}, \tilde{x}^i) = x^i$. 
Under the change of the coordinates, the line element should remain invariant, 
$
ds^2
 = - \frac{\alpha^2}{H^2 (N)} d N^2
 + a^2 (N) e^{2 \zeta} \gamma_{i j} dx^i dx^j
 = - \frac{\tilde{\alpha}^2}{H^2 (\tilde{N})} d \tilde{N}^2
 + a^2 (\tilde{N}) e^{2 \tilde{\zeta}} \tilde{\gamma}_{i j}
 d \tilde{x}^i d \tilde{x}^j$. 
Equating the coefficients of $d \tilde{N}^2$,
$d \tilde{N}d \tilde{x}^i$, and $d \tilde{x}^i d \tilde{x}^j$
on both sides of the above, we obtain the nonlinear gauge 
transformation rules in Appendix of \cite{Naruko:2012fe}. 
\subsection{Beyond $\delta N$ formalism}
Let us briefly 
summarise the five steps in the {\it Beyond $\delta N$ formalism}.

\begin{enumerate}
 \item Write down the basic equations (the Einstein equations and
 scalar field equation) in the uniform $\calN$ slicing with 
$\beta^i=0$ of (\ref{assum:beta-zero}). For convenience let us call
the choice of the coordinates in which one adopts the 
 uniform $X$ slicing with $\beta^i=0$, the $X$ gauge. 
So the above choice is the $\calN$ gauge.
In this gauge the metric components at leading order are trivial
since both $\zeta$ and $\gamma_{ij}$ are independent of time.

\item First solve the leading order scalar field equation
 under an appropriate initial condition and then the next-to-leading order
 scalar field equation which involves spatial gradients
 of the leading order solution.

\item Solve the next-to-leading order Einstein equations for 
the metric components and their derivatives.

\item Determine the gauge transformation from the $\calN$ gauge 
 to the $K$ gauge and apply the gauge transformation rules
 to obtained the solution the obtained solution in the $K$ gauge. 

 \item Evaluate the curvature perturbation $\frR = \zeta + \chi/3$
 in the $K$ gauge, where $\chi$ is to be extracted from $\gamma_{i j}$.
\end{enumerate}
\subsection{Solvable example}
\label{sec:sbrid}
In this subsection, we demonstrate how to obtain the solution up to
next-to-leading order in gradient expansion by applying our formalism
to a specific, analytically solvable model. 
We consider a canonical scalar field 
 with exponential potential~\cite{Sasaki:2008uc}, 
 \begin{align}
 P = \frac{1}{2} X_{I J} - V (\phi_I) \,, \quad 
 V (\phi_I) = W \exp \left[ \sum_J m_J \phi_J
 \right] \,,
 \end{align}
 where $W$ is a constant. 
The solution is obtained 
under the assumption of slow-roll conditions by
\begin{align}
 {}\phi^{(0)}_I (N) = C_I^\phi + m_I (N - N_0) \,,
\end{align}
 and the next-leading order solution is also obtained as
\begin{align}
 {}\phi^{(2)}_I 
 &= \frac{1}{3} D_I^\phi \Bigl[ e^{3 (N - N_0)} - 1 \Bigr]\notag\\
 &+ \int_{N_0}^N dN' e^{3 N'} \int^{N'}_{N_0} dN'' e^{- 3 N''}
 \frac{S_I^\phi}{a^2 e^{2 C^\zeta} V^{(0)}} \,,
\label{sb:sol-phi-2}
\end{align}
where
\begin{align}
 S^\phi_I 
 =& \, 3 \frac{K^{(0)}}{e^{C^\zeta}} \pa_i \left( \frac{e^{C^\zeta}}{{}K^{(0)}}
 D^i{}\phi^{(0)}_I \right)\notag\\ &+ {}K^{(0)} \left[ D^2 \left( \frac{1}{{}K^{(0)}} \right)+ D^i \left( \frac{1}{{}K^{(0)}} \right) D_i C^\zeta \right] \pa_N {}\phi^{(0)}_I
 \notag\\
 & \qquad - \Bigl[ R - \bigl( 4D^2 C^\zeta + 2 D^i C^\zeta D_i C^\zeta \bigr)
 \notag\\&\qquad + 2 D^i{}\phi^{(0)}_J D_i {}\phi^{(0)}_J \Bigr] \pa_N {}\phi^{(0)}_I \,,
\label{sb:sphi}
\end{align}
$N_0$ is an initial time and $C_I^\phi$ and $D_I^\phi$ 
 represent the initial values of the scalar field and its time derivative. 

We obtain analytic solutions for all variables 
on the uniform ${\cal N}$ gauge. 
We derive the solution on the $K$ gauge 
 by applying a gauge transformation to the solution on the $\calN$ gauge. 
To do so, we first need to determine the generator of the gauge transformation
 between the two slices, $N \to \tilde{N} = N + n (N, x^i)$ or conversely
 $\tilde{N} + \tilde{n} (\tilde{N}, \tilde{x}^i) = N$.

What we need to know is the final value of $\frR$ at sufficiently late times, 
$N\to 0\ (a\to a_0 e^{N_0})$. We take $N_0$ to be a time around which the
scales relevant to cosmological observations crossed the Hubble horizon, 
hence $N_0\gtrsim50$.
In this case, at $N=0$, the curvature perturbation reduces to
\begin{align}
\frR_K(N=0)\approx {}^{(0)}C^{\zeta}+{}^{(2)}C^{\zeta}-{m_I\over 3 M^2}D^\phi_I
\,.\label{sb:late-time-R}
\end{align}
where we have defined $M^2=\sum_I m_I^2 $ and assumed small mass: $M^2\ll 1$. 
The first term, ${}^{(0)}C^{\zeta}$ represents the leading order
curvature perturbation obtainable in the usual $\delta N$ formalism,
and the remaining terms represent $\calO(\epsilon^2)$ contributions, 
the calculation of which is the main purpose of the beyond $\delta N$ formalism. The small mass $M$ can give a contribution to large effect on the last term, 
corresponding to initial time derivative of scalar field $\phi^I$. 
\section{Beyond Einstein gravity}
In this section, we will consider gradient expansion formalism for 
a general single field in beyond GR gravity, namely modified gravity. 
Such theory is related to a general scalar-tensor theory. 
We take the Lagrangian density of beyond Horndeski (GLPV) theory 
\cite{Gleyzes:2014dya} as a boarder generalization of the Galileons to 
curved spacetime, 
which is of the form ${\cal L}=\sqrt{-g} \sum_a L^\phi_a$, with 
\begin{align}
L_2^\phi=& G_2(\phi,X)\,,\quad L_3^\phi=G_3(\phi,X)\Box 
\phi\,,\notag\\
L_4^\phi=&G_4(\phi,X)\, ^{(4)}R-2 G_{4,X}(\phi,X) (\Box\phi^2 
-\phi^{\mu\nu}\phi_{\mu\nu})\,,\notag\\
&+F_4(\phi,X)e^{\mu\nu\rho}_\sigma e^{\mu'\nu'\rho'\sigma}\phi_\mu\phi_{\mu'}
\phi_{\nu\nu'} \phi_{\rho\rho'}\,,\notag\\
L_5^\phi=&G_5(\phi,X)\, ^{(4)}G_{\mu\nu}\phi^{\mu\nu}\notag\\
&+{1\over 3} G_{5,X}(\phi,X)(\Box\phi^3-3\Box \phi \phi_{\mu\nu} 
\phi^{\mu\nu} +
2 \phi_{\mu\nu}\phi^{\mu\sigma}\phi^\nu_\sigma)\notag\\
&+F_5(\phi,X)e^{\mu\nu\rho\sigma}e^{\mu'\nu'\rho'\sigma'}\phi_{\mu}
\phi_{\mu'} \phi_{\nu\nu'}\phi_{\rho\rho'}\phi_{\sigma\sigma'}\,,
\label{Galileons}
\end{align}
where $\phi_\mu\equiv \na_\mu \phi$, 
$\phi_\nu\equiv \na_\nu \phi$, $e_{\mu\nu\rho\sigma}$ is the totally 
antisymmetric Levi-Civita tensor and 
$^{(4)}G_{\mu\nu}$ is the four-dimensional Einstein tensor. 
Note $F_4=F_5=0$ equals to Horndeski theory \cite{Horndeski}. It ensures that 
the equation of motion are second order. 
Choosing the uniform scalar field ($\phi=$const) hypersurfaces leads to a ADM 
Lagrangian of the form ${\cal L}=\sqrt{-g} \sum_a L_a$, with 
\begin{align}
L_2=&A_2(t,\alpha), ~~L_3=A_3(t,\alpha) K,\notag\\
L_4=& A_4(t,\alpha)(K^2-K_{ij}K^{ij})+
B_4(t,\alpha)\hat{R},\notag\\
L_5=&A_5(t,\alpha)(K^3-3K K_{ij}K^{ij}+2K_{ij}K^{ik}K^j_k)\notag\\
&+B_5(t,\alpha)K^{ij} \left(\hat{R}_{ij}-{1\over 2} g_{ij} \hat{R}\right)\,,
\label{GLPV-ADM}
\end{align}
where $K_{ij}$ and $\hat{R}_{ij}$ are the extrinsic and intrinsic curvature tensors for $g_{ij}$ on the constant time hypersurface. We set $K=g^{ij}K_{ij}$ and $\hat{R}=g^{ij} \hat{R}_{ij}$. 
The coefficients in (\ref{GLPV-ADM}) are related to the original functions 
in (\ref{Galileons}) as shown 
explicitly in \cite{Gleyzes:2014dya}. The linear perturbation theory was already studied in \cite{Kobayashi:2015gga}. We will adopt gradient expansion approach to this theoretical setup (see also our recent work \cite{Takamizu:2018} in the same point of view).

In the setup of the 
Lagrangian (\ref{GLPV-ADM}),  the 
Hamiltonian constraint is obtained by 
\begin{align}
&(A_2\alpha)'+A_3'\alpha K +(A_4/\alpha)'\alpha^2 \left({2\over 3}K^2-A_{ij}A^{ij}\right)\notag\\
&+(A_5/\alpha^2)'
\alpha^3 \left({2\over 9} K^3-K A_{ij}A^{ij} +2 A_{ia}A^a_{j}A^{ij}\right)\notag\\
&+(B_4 \alpha)' \hat{R}-{\alpha\over 6}{B}_5'K\hat{R}+\alpha B_5'\hat{R}_{ij}A^{ij}=0\,,
\end{align}
where  a prime represents differentiation with respect to $\alpha$. The 
Momentum constraint is obtained by 
\begin{align}
&D^i A_3+{4\over 3} D^i( A_4 K)-2 e^{-3\zeta}
D_j(A_4 e^{3\zeta} A^{ij})\notag\\&+D^i
\left(A_5 \left({2\over 3}K^2 -3 A^{ab}A_{ab}\right)\right)\notag\\
&-2 e^{-3\zeta}
D_j \left(A_5 e^{3\zeta}(K A^{ij}-3 A^{ia}A_a^j)\right)\notag\\&+
\hat{R}^{ij} D_j B_5 -{1\over 2}  D^i B_5 \hat{R}=0\,,
\end{align}
where  $D$ represents covariant derivative for $\gamma_{ij}$. 

Two dynamical equations for $(K, A_{ij})$ can be obtained by varying the Lagrangian with respect to $g_{ij}$, 
that corresponds to trace part and traceless part as 
\begin{widetext}
\begin{align}
(2A_4+2A_5 K)\,\partial_\perp K&-3 A_5 A^{ij}\partial_\perp A_{ij} 
-{3\over 2} A_2 +{3\over 2}\partial_\perp A_3 +2 K \partial_\perp A_4 
+\left(K^2-{3\over 2} A_{ij}A^{ij}\right) \partial_\perp A_5 \notag\\
&+\left(K^2+{3\over 2} A_{ij}A^{ij}\right)A_4+\left({2\over 3} K^3+3 A_{ia}A^a_jA^{ij}\right)A_5\notag\\&-{B_4\over 2} \hat{R}+{2\over \alpha} D^2(\alpha B_4)-{1\over 4}\partial_\perp {B}_5\, \hat{R}+{1 \over 2} D_i D_j B_5 A^{ij}-{1\over 3}K (D^2 B_5) -{2\over 3} D^i K D_i B_5\notag\\&+
D_i A^{ij} D_j B_5-{2\over 3\alpha} D^i \alpha D_i B_5 \, K+{1\over \alpha}
D_i \alpha A^{ij} D_j B_5=0\,,\label{eq:fullevolution-K}\\
-(A_4+A_5 K)\,\partial_\perp A_{ij}& +6 A_5 (\partial_\perp A_{ia} A^a_j)^{TF}-A_5\partial_\perp K A_{ij} -K(A_4 +A_5 K)A_{ij}\notag\\
&+{2}(A_4+A_5K)A_{ia} A^{a}_j +3 A_5 K(A_{ia}A^a_j)^{TF}-12 A_5 (A_{ib} 
A^b_a A^a_j)^{TF}\notag\\
&-\partial_\perp A_4 A_{ij} +(\partial_\perp A_5) \left(-KA_{ij}+3 (A_{ia}A^a_j)^{TF}\right)\notag\\
&+B_4 \hat{R}_{ij}^{TF}-{1\over \alpha}[D_i D_j(\alpha B_4)]^{TF}+{1\over 2}\partial_\perp B_5 \,\hat{R}^{TF}_{ij}+{K\over 6}(D_i D_j B_5)^{TF}+{1\over 2}(D^2 B_5) A_{ij}\notag\\&-(A^a_i D_j D_a B)^{TF}-(D_a A^a_i D_j B_5)^{TF}+D^a A_{ij}D_a B_5+{1\over 3}(D_i B_5 D_j \, K)^{TF} -(D_i A^a_j D_a B_5)^{TF} 
\notag\\
&+{1\over \alpha} D^a \alpha D_a B_5 A_{ij}-{1\over \alpha} (D_a \alpha D_i B_5 A^a_{j})^{TF}-{1\over \alpha} (D_i \alpha D_a B_5 A^a_{j})^{TF}+{K\over 3\alpha} (D_i \alpha D_j B_5 )^{TF}=0\,,
\label{eq:fullevolution-A}
\end{align}
\end{widetext}
We will solve a general solution valid up to $\calO(\epsilon^2)$ later. 
\subsection{Background equations}
We obtain the background part of the Lagrangian as 
\begin{align}
{ \cal L}^{(0)} =
a^3 (A_2\alpha+ 3 A_3 H +6A_4 H^2/\alpha +6A_5 H^3/\alpha^2)\,,
\label{background-L}
 \end{align} 
by using  $K=3H/\alpha$, where a bar means a background quantity.
Varying Eq. (\ref{background-L}) with respect to 
$\alpha$ and $a$, we obtain, respectively,
\begin{align}
-{\cal E}:=\ &(A_2 \alpha)'+3 A_3' H+6 (A_4/\alpha)' H^2 \notag\\
&+6 (A_5/\alpha^2)' H^3=0\,,
\label{eq:background1}\\
{\cal P}:=\  & A_2\alpha -6A_4 H^2/\alpha-12 A_5 H^3/\alpha^2\notag\\
&-{d\over dt}(A_3+4 A_4 H/\alpha+6 A_5 H^2/\alpha^2)=0\label{eq:background2}\,.
\end{align}
\subsection{Leading-order}
The leading-order in the gradient expansion can be interpreted as 
$\delta N$ for curvature perturbation. 
In this representation, 
inhomogeneous parts can be interpreted as perturbed quantities $\zeta$
\begin{align}
&\zeta(t,\bm x)=\delta N:={\cal N}-N(t)\,,
\label{eq:hij-deltaM}
\end{align}
where $N$ represent a background $e$-folding number. 
If one takes FLRW background, i.e. $a=e^N(t)$, $\zeta$ quantifies 
the curvature perturbation as $\delta N$ in the view of 
the separate Universe approach. That is the so-called $\delta N$ formalism. 

In this approach, trace and traceless parts of the extrinsic curvature can 
be given by
\begin{align}
K=3 {dN\over dt}+{\cal O}(\epsilon^2)\,,~~~A_{ij}={1\over 2} {d\gamma_{ij}\over dt}+{\cal O}(\epsilon^2)\,.
\end{align}
Note that $A_{ij}$ represents a cosmic shear rate. 

One can obtain ADM equations (\ref{eq:fullevolution-K}) and 
(\ref{eq:fullevolution-A}) in the leading-order as 
\begin{align}
2\Xi\, \partial_t K= &{\cal G}_K\,K -3A_5\alpha A^{ij}\partial_t A_{ij}\notag\\
&+{3\over 2}(A_4\alpha-\dot{A}_5)
{A_{ij}A^{ij}}\notag\\
&+3A_{ik}A^k_j A^{ij} A_5\alpha +{\cal O}(\epsilon^2)\,,
\label{eq:evol-K-leading}\\
\Xi\,\partial_t A_{ij}= &-(3H\Xi+\partial_t\Xi )A_{ij}-6A_5\alpha \partial_t A_{ia}A^a_j
\notag\\
&-(2A_4\alpha +15A_5H+3\dot{A}_5) A_{ik} A^k_j\notag\\
&+6A_5\alpha A_{il}A^l_k A^k_j+{\cal O}(\epsilon^2)\,.
\label{eq:evol-A-leading}
\end{align}
where a dot represents differentiation with respect to $t$ and 
we defined 
\begin{align}
\Xi(t)=\,&-(A_4+3 A_5H/\alpha )\,,\\
{\cal G}_K(t)=\,&3H \Xi+\dot{A}_4+3\dot{A}_5H/\alpha   \,.
\end{align}
If one consider the GR case, it takes $A_4=-B_4=-1/2$, $A_2=P(\phi,X)$ and all others vanishing. So the definition of $\Xi$ is one as being a positive value. Here we assume that cosmic shear is weak, i.e. $A_{ij}\ll 1$. It can allow us to ignore the quadratic and cubic terms: $(A_{ij}^2,A_{ij}^3)$. Then (\ref{eq:evol-A-leading}) reads 
\begin{align}
\partial_t A_{ij}\simeq  -\{3H+\partial_t( \ln \Xi)\} A_{ij}+{\cal O}(A_{ij}^2,\epsilon^2)\,,
\end{align}
and it shows a leading order solution of $A_{ij}$ can be ignored since 
it is just a decaying mode, i.e. $A^{(0)}_{ij}\propto a^{-3}\Xi^{-1}$ in the 
context of inflationary universe with $a\to \infty$. 
So we can set for general situation 
\begin{align}
A_{ij}={\cal O}(\epsilon^2)\,,
\label{cond:Aij}
\end{align}
that is same condition for (\ref{stronger condition for gamma}), 
which we have used in our procedure. 
Then (\ref{eq:evol-K-leading}) also can be simplified as 
\begin{align}
\partial_t K\simeq \{{\cal G}_K/(2\Xi)\} K+{\cal O}(A_{ij}^2,\epsilon^2)\,,
\end{align}
If it shows ${\cal G}_K(t)/\Xi<0$, the perturbation of $K$ also can be 
ignored since it just decays in time as $K\propto \exp\left[\int {\cal G}_K/(2\Xi) dt\right]$ in an inflationary cosmology
\footnote{The exact solution of the perturbation of $K$ can be 
obtained as (\ref{solution:delta-K}) shown in the later subsection. It is shown as a solution at the next-leading order, but it is same solution at the leading order whenever the perturbations of $K$ and $\alpha$ can be dealt with 
as small quantities, that is solution for linearized equations. The exact condition for decaying mode in this limit, ${\cal G}_D<0$ in (\ref{def:calG-d})}.

\subsection{Next-leading order}
In this subsection, we will adopt the condition (\ref{cond:Aij}). 
We can introduce the perturbations of $K$ and $\alpha$ as 
\begin{align}
&K={3H(t)\over \alpha(t)}\left[1+\delta K(t,x^i)\right]\notag\\
&\alpha=\alpha(t)\left[1+\delta \alpha(t,x^i)\right]\,.
\end{align}
The condition (\ref{cond:Aij}) can read 
\begin{align}
\partial_t \gamma_{ij}={\cal O}(\epsilon^2)\,,
\label{cond:gamma}
\end{align}
that is equivalent to recovering FLRW Universe in the limit of taking $\epsilon\to 0$ because a shear of the Universe vanishes at the leading order. 
As shown in the last subsection, we can also take 
\begin{align}
\delta K=\delta \alpha ={\cal O}(\epsilon^2)\,.
\end{align}
Up to the order of ${\cal O} (\epsilon^2)$, we can obtain ADM equations at the next-leading order, that are Hamiltonian constraint and two dynamical equations (\ref{eq:fullevolution-K}) and 
(\ref{eq:fullevolution-A})  as 
\begin{align}
&\delta \alpha=-{ {\cal G}_A \over \Lambda}\,\delta K
-{{\cal F}_A\over \Lambda}\, \hat{R}+{\cal O}(\epsilon^4)\,,
\label{evol:alpa}\\
&\Xi \, \partial_t (\delta K)-\lambda \partial_t (\delta \alpha)=
 { {\cal G}_B }\,\delta K+{{\cal G}_C}
\,\delta \alpha\notag\\
&\quad \quad \quad -{\alpha^2{\cal F}_B\over 12H} \,
\hat{R}+{\cal O}(\epsilon^4)\,,
\label{evol:dK}\\
&\partial_t A_{ij}= {{\cal F}_C\over \Xi} \, A_{ij}-{\alpha{\cal F}_B\over 
a^{2}e^{2\zeta}\,\Xi}\,  [\hat{R}_{ij}]^{\rm TF}+{\cal O}(\epsilon^4)\,,
\label{evol:Aij}
\end{align}
where the coefficients are defined as 
\begin{align}
\Lambda= & \,\alpha({A}_2\alpha)''+
3 H ({A}_3'\alpha)'\notag\\
&+6 H^2{A}_4''+6 H^3 ({A}_5 '/\alpha)'\,,\\
{\cal G}_A= &\, 3 H A_3'+12 H^2 (A_4/\alpha)'+18 H^3 (A_5/\alpha^2)'\,,\\
{\cal F}_A= &\,(B_4\alpha )'/\alpha-B_5'H/2\,,\\
{\cal G}_B= &\, -3 H\Xi -\partial_t\ln \left({H/\alpha}\right)\Xi
\notag\\&+ (\dot{A}_4+
3H\dot{A}_5/\alpha) \,,\\
\lambda=&\, \alpha^2A_3'/4H +A_4'\alpha+3HA_5'/2\,,\\
{\cal G}_C= & \,-{\alpha (A_2\alpha)'\over 4H}+{3H\over 2\alpha}  ({A}_4 \alpha)'+{3H^2 \over \alpha}({A}_5\alpha)'\notag\\
&+{\alpha\over 4 H}\partial_t (A_3'\alpha)+\partial_t (A_4'\alpha)+
{3H\over 2\alpha} \partial_t (A_5'\alpha)\,,\\
{\cal F}_B= & \,B_4+{\dot{B}_5/ (2\alpha)}\,,\\
{\cal F}_C= & \,-3H\Xi-\pa_t \Xi \,,
\end{align}
and we have used the fact that 
any spatial derivative terms of $B_4$ and $B_5$ can be estimated as 
higher expansion order, that is a same order of $D_i D_j \delta \alpha={\cal O}(\epsilon^4)$ via $B_4(t,\alpha(t,x^i))$ with the covariant derivative $D$ for the metric $\gamma_{ij}$. Note that (\ref{evol:Aij}) is a linearized 
equation for $A_{ij}$ since we assumed cosmic shear is weak with 
(\ref{cond:Aij}).

Ricci scalar and Ricci tensor can be rewritten as 
\begin{align}
\hat{R}=&a^{-2} e^{-2\zeta} \left[ 
R-(4D^2 \zeta +2 D_i\zeta D^i \zeta) 
\right]\,,\label{hat_R}\\
\hat{R}_{ij}^{\rm TF}=&[R_{ij}+D_i \zeta D_j\zeta-D_i D_j \zeta]^{\rm TF}\,.
\label{eq:Riccitensor-gamma}
\end{align}

If  one focus on a scalar mode, 
it needs to solve two eqs (\ref{evol:alpa}) and (\ref{evol:dK}) for $\delta \alpha$ and  $\delta K$, respectively.  When the coefficient $\lambda$ is non-zero, their time 
evolution equations are coupled, but they are reduced to one equation by substituting (\ref{evol:alpa}) into (\ref{evol:dK}) as
\begin{align}
\partial_t (\delta K)= { {\cal G}_D }\,\delta K-{\cal F}_D\,
(a^2\hat{R})+{\cal O}(\epsilon^4)
\end{align}
The solution of $\delta K$ can be obtained by
\begin{align}
\delta K=G(t) \left(C^{(2)} -(a^2\hat{R})\int^t {\cal F}_D(t') /G(t') \,dt' \right)\,,\label{solution:delta-K}
\end{align}
where $C^{(2)}$ is an integration constant and we defined 
\begin{align}
G(t)=&\,\exp\left[{\int {\cal G}_D(t) dt}\right]\,,\\
{\cal G}_D(t)=&\,\Biggl[{{\cal G}_B}-{{\cal G}_A{\cal G}_C
\over \Lambda }-\lambda \partial_t \left({{\cal G}_A\over \Lambda}\right)\Biggr]/L\,,\label{def:calG-d}\\
{\cal F}_D(t)=&\,\Biggl[{{\cal F}_A {\cal G}_C\over \Lambda a^2}+
{\alpha^2 {\cal F}_B\over 12H a^2 }
+\lambda \partial_t \left({{\cal F}_A\over \Lambda a^2}\right)\Biggr]/L\,,\\
L=&\ \Xi+\lambda\,  {\cal G}_A/ \Lambda \,.
\end{align}
Note that the term $a^2 \hat{R}$ is a just spatial function as shown in 
(\ref{hat_R}). The term ${\cal G}_C$ can be rewritten by using background equations (\ref{eq:background1}) and (\ref{eq:background2}) as
\begin{align}
&{\cal G}_C=\,-{\alpha \over 4}(3 A_3'  -6 A_5' (H/\alpha)^2))
\notag\\
&+{\alpha\over 4H}(\partial_t(A_3'\alpha)+4H/\alpha \partial_t(A_4'\alpha) +
6(H/\alpha)^2 \partial_t (A_5' \alpha))\,.
\end{align}
In the GR limit, it can show easily that the term ${\cal G}_C=0$ and 
also $\lambda=0$. 
In this case, one can also show ${\cal G}_D={\cal G}_B/\Xi=-3H-
\partial_t\ln (H/\alpha)$ and it can simplify the solution of $G(t)$ as 
$a^{-3}H/\alpha$, that is a decaying mode with inflation. Note that 
this simple solution is related to a discussion in 
the previous subsection about leading-order.

Therefore a curvature perturbation can be finally 
obtained as 
\begin{align}
\partial_t \zeta =\,H(\delta K+\delta \alpha)
+{\cal O}(\epsilon^4)\,.
\end{align}
The final solution of $\zeta$ can be obtained by
\begin{align}
\zeta=\left[\int^t H\,G(t')\left(1-{{\cal G}_A\over \Lambda}\right)dt'\right]\,C^{(2)}-
{\cal F}_R(t)\,(a^2\hat{R})\,,
\end{align}
with 
\begin{align}
{\cal F}_R(t)=&\int^t ( H {\cal F}_A/\Lambda)\,dt'\notag\\&+\int^t 
\left[H\,G(t')\left(1-{{\cal G}_A\over \Lambda}\right)\int^{t'}({\cal F}_D/G) dt''\right]dt'.
\end{align}
This is our main result for curvature perturbation in general scalar-tensor theory. However, it is impossible to construct one master 
evolution equation since this solution is related to 
two contributions: $\delta \alpha$ and $\delta K$, which is different from 
the case of (\ref{pa_zeta-alpha}). So let us see 
tensor mode in detail in the next subsection. 
\subsection{Gravitational waves via nonlinear interactions}
We focus on the tensor mode in this subsection. 
Tensor perturbation can be easily dealt with because of no coupling with 
$\delta K$ and $\delta \alpha$ up to the next-leading order in the 
gradient expansion. 
First, we obtain the general solution under the condition: (\ref{cond:gamma}) by integrating a basic equation for $A_{ij}$: (\ref{evol:Aij}), 
\begin{align}
A_{ij}= e^{\int \Theta dt }\left\{C^{(2)}_{ij}-{\cal F}^{(2)}_{ij}
\int \left( {\alpha c_T^2\over a^2}e^{-\int \Theta dt}\right)dt 
\right\}\,,
\end{align}
with 
a sound speed squared: $c_T^2$; 
\begin{align}
c_T^2={{\cal F}_B\over \Xi}={2B_4\alpha 
+\partial_t B_5 \over-( 2A_4\alpha +6 H A_5)}\,,
\label{sound speed of GW}
\end{align}
and we defined two new quantities:  
\begin{align}
\Theta={{\cal F}_C \over \Xi}=-3 H-\pa_t (\ln \Xi)\,,
\label{eq:Theta}
\end{align}
and 
\begin{align}
{\cal F}^{(2)}_{ij}(\bm x)={\hat{R}_{ij}^{\rm TF}[\gamma,\zeta]\over e^{2\zeta}}\,,
\label{source-interaction}
\end{align}
and integral constant: $C^{(2)}_{ij}$. 
When one can use this solution, it is easy to integrate 
(\ref{basic-gamma}) and reads 
\begin{align}
\gamma_{ij}=\gamma_{ij}^{(0)}+D(t)\, C^{(2)}_{ij}+ F(t)\,  {\cal F}^{(2)}_{ij}
+\calO(\epsilon^4)\,,  
\label{eq:gamma-sol}
\end{align}
where 
\begin{align}
D(\tau)=
& \,2 \int^\tau {\alpha(\tau')\over z^2} \,d\tau' \,,\label{D-decayingGW}\\
F(t)=&\,-2\int^\tau {\alpha(\tau')\over z^2}\, \Bigl(\int^{\tau'} 
 {\alpha c_T^2 z^2 (\tau'') } d\tau'' \Bigr)\,d\tau'\,,
\label{F-growingGW}
\end{align}
with 
\begin{align}
z:=a\sqrt{\Xi}=a\sqrt{-(A_4+3 A_5 H/\alpha)}\,,
\label{z-GW}
\end{align}
where we used $e^{\int  \Theta dt}=1/(a z^2) $ from (\ref{eq:Theta}). 
Here $\gamma_{ij}^{(0)}$ represents some spatial function as an 
integral constant at the leading-order. 
The integral $D(t)$ and $F(t)$ are related to decaying and growing modes, respectively. In the GR limit, we have $z=a$. So the integral results 
$D(t)\propto \int a^{-3} dt$, that 
shows decaying mode. In this case, the integral 
$\int z^2 d\tau  $ can be reduced to $\int a dt $, corresponding to a growing mode. Therefore, the integrals tell us to give a possibility that tensor perturbation can be enhanced 
when $z$ decreases or $z^2 c_T^2$ increases in $\tau$.

Next, we can easily 
obtain one master evolution equation for tensor perturbation 
$\gamma_{ij}$ 
by using the solution of (\ref{eq:gamma-sol}) as 
\begin{align}
{\partial^2 \gamma_{ij}\over \partial \tau^2}+
2 {\pa_\tau z\over z}\,{\partial \gamma_{ij}\over \partial \tau}+2 c_T^2{\cal F}^{(2)}_{ij}[\gamma]={\cal O}(\epsilon^4)\,.
\label{evolution eq for gamma GW}
\end{align}
Here the meaning 
of the term $c_T^2$ defined in (\ref{sound speed of GW}) denotes 
some propagation speed of gravitational waves. Note that the 
GR limit always gives constant sound speed $c_T=1$ with $A_4=-B_4=-1/2$ and 
the functions $A_5$ and $B_5$ can allow us to change $c_T^2$. 
The value of (\ref{sound speed of GW}) is 
consistent with the linear perturbation 
theory of GLPV theory in \cite{Kobayashi:2015gga}. If one consider the time varying $B_5$, the speed of gravitational waves can change in time. 
It is related to enhancement of 
gravitational waves on superhorizon scale via time evolution on the effect of 
${\cal O}(\epsilon^2)$. And also the variable $z$ of (\ref{z-GW}) can lead to 
the condition of the enhancement of 

(\ref{evolution eq for gamma GW}) is compatible with 
linear equation \cite{Kobayashi:2015gga}: 
\begin{align}
{\partial^2 \gamma_{ij}\over \partial \tau^2}+2 {\pa_\tau z\over z}\,
{\partial \gamma_{ij}\over \partial \tau}-c_T^2 \Delta \gamma_{ij}=0\,.
\end{align}
The source term (\ref{source-interaction}) 
is a given function only depending on spatial coordinate via 
(\ref{eq:Riccitensor-gamma}). 
This term is source term for a time evolution of $\gamma_{ij}$ and 
includes all full-nonlinear interaction over scalar and tensor modes. 
Actually, when you expand (\ref{source-interaction}) as any interaction via 
$(\gamma_{ij} \times \gamma_{ij}), (\gamma_{ij}\times \zeta), (\zeta\times \zeta),\dots, (\zeta\times \zeta\times \gamma_{ij})\,, etc$. That gives 
a main difference between linear and nonlinear theory. 
It maybe affect non-Gaussianity of tensor modes. 
Of course, (\ref{evolution eq for gamma GW}) can be reduced to 
a usual GR linear theory as 
\begin{align}
{\partial^2 \gamma_{ij}\over \partial \tau^2}+2 {\cal H}
{\partial \gamma_{ij}\over \partial \tau}-\Delta \gamma_{ij}=0\,,
\end{align}
where the last source term can be derived from $\hat{R}_{ij}[\gamma]=-\Delta 
\gamma_{ij}/2+\calO(\gamma^2)$ in (\ref{eq:Riccitensor-gamma}) with $c_T^2=1$ and $z=a$ in the GR limit. 

This basic equation (\ref{evolution eq for gamma GW}) is a main result for gravitational mode since it contains 
tensor mode plus scalar mode $\chi$ defined by (\ref{def: nonlinear HT1}). 
The quantity $\chi$ is important for construct nonlinear 
curvature perturbation: ${\frR}^{\rm NL}$ in 
(\ref{def0: nonlinear variable zeta1}), whose contribution of the 
next-leading order in expansion is given by $\chi=\calO(\epsilon^2)$. 
 In the GR limit, the evolution equation (\ref{evolution eq for gamma GW}) 
can be reduced to 
(\ref{evolution gamma GR}) via $2 \pa_\tau \ln z=2{\cal H}$ and $c_T^2=1$. 

On the other hand, 
the tensor perturbation can be extracted from $\gamma_{ij}$ by 
taking a perturbation for $h_{ij}=\gamma_{ij}-\delta_{ij}$ in 
the weak limit of $h_{ij}$ and using the following decomposition; 
where all symmetric traceless tensors: $X_{ij}$ can be decomposed as 
\begin{align}
X_{ij}=&{3\over 2} (\tilde{k}_i\tilde{k}_j-{1\over 3} \gamma_{ij})X_{\parallel}+2\sum_{a=2,3} \tilde{k}_{(i} e^a_{j)} X_a\notag\\
&+\sum_{\lambda=+,\times} e^\lambda_{ij}X_\lambda\,,
\end{align}
where $k_i$ denotes a constant comoving co-vector 
since standard plane waves are used as basis at each constant time hypersurface, while the direction of wave vector: $k^i$ changes with time. Here we defined  
the unit vector 
$\tilde{k}=k^i/\sqrt{k_ik^i}$, orthogonal basis set spanning the constant time hypersurface: 
$(\tilde{k}_i,e^2_i, e^3_i)$,  and 
a polarization tensor $e^\lambda_{ij}$; defined as $
e^\lambda_{ij}={(e^2_ie^2_j-e^3_ie^3_j )}\delta^\lambda_+ /\sqrt{2} +
{(e^2_ie^3_j+e^3_ie^2_j)}\delta^\lambda_\times/\sqrt{2}$. 
By using this method, 
the tensor perturbation $h_{ij}$ also can be decomposed into $(h_\parallel, h_a, h_{+,\times})$. The physical tensor perturbations, namely gravitational wave 
modes are $h_{+,\times}$ in the decomposition. 
\section{Summary and discussion}
\label{sec:summary}

In this paper, we developed a theory of nonlinear cosmological
 perturbations on superhorizon scales in the context of inflationary 
cosmology. First, we followed GR gravity plus a general kinetic 
single inflaton. In this case, the energy-momentum tensor for the scalar field is equivalent to that of a perfect fluid. We have solved the field equations using spatial gradient expansion in terms of a small parameter $\epsilon=k/(aH)$, where $k$ 
is a wavenumber, and obtained
a general solution for the metric and the scalar field up to $\calO(\epsilon^2)$. Then we show a matching condition between nonlinear solution and linear solution, but including $k^2$ effect. The master evolution equation is a key 
result to characterize this system compatible to similar evolution equation in 
linear perturbation theory.

Second, we extend this formalism to apply Galileon-inflation, for which
the inflaton Lagrangian is added by $G(X,\phi)\Box\phi$. 
In the case of G-inflation, it can no longer be recast into 
a perfect fluid form,
and hence its imperfect nature shows up
when the inhomogeneity of the Universe is considered. 
We have solved the field equations using spatial gradient expansion 
and also obtained
a general solution up to $\calO(\epsilon^2)$. Then we 
introduce an appropriately defined variable for 
the nonlinear curvature perturbation in the uniform $\phi$ (comoving) gauge, 
${\frR}_c^{\rm NL}$. Upon linearization, this variable
reduces to the previously defined linear curvature perturbation
${\cal R}_c^{\rm Lin}$ on a comoving hypersurfaces.
It has been also shown that ${\frR}_c^{\rm NL}$
satisfies a nonlinear second-order differential equation~(\ref{eq: basic eq for NL}), which is a natural extension of both linear perturbation equation for ${\cal R}_c^{\rm Lin}$ and nonlinear for ${\cal R}_c^{\rm NL}$ shown in 
(\ref{O29-05_eq: basic eq for NL}). 
We have shown some applications of our formalism and 
the effect is sensitive to the temporal violation of 
some kind of slow-roll conditions. So our formalism is good to evaluate 
time evolution of curvature perturbation in the physics for violation of slow-roll conditions. 

We considered a multi-component scalar field
 with a general kinetic term and a general form of the potential.
To discuss the superhorizon dynamics, we employed the ADM formalism
 and the spatial gradient expansion approach. 
Different from the single-field case, there is a difficulty in 
solving the equations in the multi-field case.
At leading-order, the equations take the same form as
those for the homogeneous and isotropic FLRW background
with suitable identifications of variables.

In cosmological perturbation theory, the most important quantity
to be evaluated is the curvature perturbation on the comoving
 slices which is conserved on superhorizon scales after the universe 
has reached the adiabatic limit. This quantity accurate to next-to-leading
order may be relatively easily obtained in the single-field case because
of the above mentioned coincidence among several temporal slicings. 
On the other hand, in the multi-field case, such a coincidence 
between different slicings does not hold. 
We first solve the field equations in a slicing in which
the lapse function is trivial. We adopt the uniform $e$-folding number slicing
in which the time slices are chosen in such a way that
the number of $e$-folds along each orbit orthogonal to the time
slices, $\calN$, is spatially homogeneous on each time slice. Then 
we can solve the equations to next-to-leading order
without encountering the above mentioned problem.
After the solution to next-to-leading order is obtained, we
transform it to the one in the uniform expansion slicing
which is known to be identical to the comoving slicing on
superhorizon scales in the adiabatic limit.
Thus the gauge transformation laws play an essential role
in our formalism. We derived them which are accurate to
next-to-leading order.
Note that they are fully nonlinear in nature in the language 
of the standard perturbation approach. 

Finally, we show an extension our formalism to beyond Einstein gravity, that is general scalar-tensor theory which can lead to several kinds of 
modified gravity. These theories are motivated not only inflation, but 
also the topic of dark energy. We used beyond Horndeski (GLPV) theory at the uniform $\phi$ gauge that includes the Horndeski theory, equivalently the 
most general second-order in the equation of motions. We construct a dynamical equation for superhorizon 
tensor perturbation with a full nonlinear interaction between scalar 
and tensor perturbation (\ref{evolution eq for gamma GW}) with $z=a\sqrt{-(A_4+3 A_5 H/\alpha)}$ as (\ref{z-GW}).  The GR case of $A_4=-1/2, A_5=0$ gives us 
$z=a$, but in general case of modified gravity n this case, $z$ can change depending on the models, namely a time varying of $A_4$ and $A_5$. The integral $\int z^{-2} d\tau$ in (\ref{D-decayingGW}) 
and $\int z^2 c_T^2 d \tau$ in (\ref{F-growingGW}) corresponding to a decaying 
and growing mode, respectively. Therefore, the integrals tell us to give a possibility that tensor perturbation can be enhanced 
when $z$ decreases or $z^2 c_T^2$ increases in $\tau$. 
More application to calculate non-Gaussianity of 
gravitational waves will be  a future task.

\acknowledgments
YT is grateful to Tsutomu Kobayashi for valuable comments and fruitful discussions on this work and also wish to acknowledge financial support by CCS, University of Tsukuba and Rikkyo University where a part of this work had been started.


\end{document}